\documentclass[twocolumn,superscriptaddress,preprintnumbers,nofootinbib]{revtex4}
\usepackage{amsmath,amssymb,longtable}
\usepackage{graphicx}
\usepackage{epsfig,color}
\usepackage{bm}
\usepackage{verbatim}
\usepackage{soul}

\begin{document}

\title{$\,$\\[-7ex]\hspace*{\fill}{\small{\emph{Preprint no}. NJU-INP 009/19}}\\[1ex] Spectrum of fully-heavy tetraquarks from a diquark+antidiquark perspective}

\author{M. A. Bedolla}\email[]{marco.bedolla@unach.mx}
\affiliation{Mesoamerican Centre for Theoretical Physics, Universidad Aut\'onoma de Chiapas, \\ Carretera Zapata Km. 4, Real del Bosque (Ter\'an), Tuxtla Guti\'errez 29040, Chiapas, M\'exico}
\affiliation{Istituto Nazionale di Fisica Nucleare (INFN), Sezione di Genova, Via Dodecaneso 33, 16146 Genova, Italy}
\affiliation{Instituto de F\'isica y Matem\'aticas, Universidad Michoacana de San Nicol\'as de Hidalgo, Edificio C-3, Ciudad Universitaria, Morelia, Michoac\'an 58040, M\'exico}

\author{J. Ferretti}\email[]{jferrett@jyu.fi}
\affiliation{Center for Theoretical Physics, Sloane Physics Laboratory, Yale University, New Haven, Connecticut 06520-8120, USA}
\affiliation{Department of Physics, University of Jyv\"askyl\"a, P.O. Box 35, 40014 Jyv\"askyl\"a, Finland}

\author{C. D. Roberts}\email[]{cdroberts@nju.edu.cn}
\affiliation{School of Physics, Nanjing University, Nanjing, Jiangsu 210093, China}
\affiliation{Institute for Nonperturbative Physics, Nanjing University, Nanjing, Jiangsu 210093, China}

\author{E. Santopinto}\email[corresponding author: ]{elena.santopinto@ge.infn.it}
\affiliation{Istituto Nazionale di Fisica Nucleare (INFN), Sezione di Genova, Via Dodecaneso 33, 16146 Genova, Italy}

\begin{abstract}
Using a relativized diquark model Hamiltonian, we calculate the masses of $J^{PC}=0^{++}$ ground-state tetraquarks in the following systems: $b s \bar b \bar s$, $bb \bar n \bar n$ ($n=u, d$), $bb \bar s \bar s$, $cc\bar c \bar c$, $b b \bar b \bar b$, $b c\bar b  \bar c$ and $b b \bar c \bar c$.
We also compute extensive spectra for the fully-heavy quark flavour combinations. 
Finally, as a test of the diquark model approach, we compute the masses of fully-heavy baryons in the diquark model. Our results may be compared soon to the forthcoming experimental data for fully-heavy three-quark systems.
\end{abstract}

\maketitle

\section{Introduction}
Until the current millennium, the spectrum of detected hadrons was limited to systems that fit simply into the patterns typical of constituent-quark models \cite{GellMann:1964nj, Zweig:1981pd}, \emph{i.e}.\ quark-antiquark $(q \bar q)$ mesons and three-quark $(qqq)$ baryons.  Notwithstanding this, Refs.\,\cite{GellMann:1964nj, Zweig:1981pd} also raised the possibility of complicated hadrons, \emph{e.g}.\ $qq \bar q\bar q$ and $q\bar q qqq$.
A few years later, Lichtenberg suggested that ``{\it exotic mesons might be realized by replacing the quark and antiquark of a normal meson with a diquark and antidiquark, respectively}'' \cite{Lichtenberg:1981pp,Lichtenberg:1978sj}. Investigations on the phenomenology of $qq \bar q \bar q$ mesons were conducted in the 70s, when Jaffe proposed a four quark interpretation for light scalar mesons \cite{Jaffe:1976ig,Jaffe:1976ih} and Iwasaki the possible existence of hidden-charm tetraquarks \cite{Iwasaki:1975pv}. Later, a similar study related to multiquark systems indicated the existence of baryon-antibaryon mesons \cite{Montanet:1980te}. Potential models to study $qq\bar q \bar q$ systems were presented in Refs. \cite{Chao:1980dv,Weinstein:1983gd}, and fully-heavy tetraquarks were hypothesized in Ref. \cite{Heller:1985cb}.

Today, a large amount of data, obtained at both $e^+ e^-$ and hadron colliders, has provided evidence for the possible existence of such \emph{exotic} hadrons.
In particular, we need to mention a recent LHCb study \cite{Aaij:2020fnh}, where the $J/\psi$-pair invariant mass spectrum was studied by using $pp$ collision data \cite{Aaij:2020fnh}. 
A narrow structure, dubbed the $X(6900)$, which matches the lineshape of a resonance, and a broad one, next to the di-$J/\psi$ mass threshold, were found. The global significance of either the broad or the $X(6900)$ structure was determined to be larger than five standard deviations.
It was stated that ``{\it The $X(6900)$ structure could originate from a hadron state consisting of four charm quarks, $T_{cc \bar c \bar c}$, predicted in various tetraquark models.}'' \cite{Aaij:2020fnh}.

Herein we focus on those exotic systems which may be considered tetraquarks, \emph{viz}.\ systems with meson-like quantum numbers that can be built using two valence quarks and two valence antiquarks.  
Some tetraquark candidates cannot be described using typical constituent quark models \cite{Godfrey:1985xj,Eichten:1978tg, Pennington:2007xr, Ortega:2009hj, Ortega:2012rs, Ferretti:2012zz, Ferretti:2013faa, Ferretti:2013vua, Ferretti:2014xqa, Ferretti:2018tco} because they carry electric charge; hence, cannot simply be $q \bar q$ systems.  Consequently, they are good candidates for:
hidden-charm/bottom tetraquarks \cite{Jaffe:1976ig, Weinstein:1983gd, Brink:1998as, Maiani:2004vq, Barnea:2006sd, Santopinto:2006my, Deng:2014gqa, Zhao:2014qva, Bicudo:2015vta, Lu:2016cwr, Eichten:2017ffp, Karliner:2017qjm, Anwar:2017toa, Hughes:2017xie, Anwar:2018sol, Wu:2018xdi, Wang:2018qpe, Chen:2019dvd, Wang:2019rdo, Ferretti:2020ewe, Richard:2019cmi},
or molecular systems constituted from a pair of charm/bottom mesons \cite{Weinstein:1990gu, Manohar:1992nd, Tornqvist:1993ng, Swanson:2003tb, Hanhart:2007yq, Thomas:2008ja};
or hadro-charmonia \cite{Dubynskiy:2008mq, Panteleeva:2018ijz, Ferretti:2018kzy, Anwar:2018bpu}.

Owing to the large masses of the valence degrees of freedom, the possible existence of fully-heavy $QQ \bar Q \bar Q$ states ($Q=c,b$) and similar, mixed systems ($c\bar c b\bar b$, $\bar c \bar c b b \sim c c \bar b \bar b$) can reasonably be explored using nonrelativistic tools for QCD  phenomenology and theory.  Here, in contrast to systems involving light-quarks, for which both light-meson and gluon exchange may play a role in tetraquark formation, binding in fully-heavy systems is very probably dominated by gluon-exchange forces because the typical gluon mass-scale ($m_g \sim 0.5\,$GeV \cite{Aguilar:2015bud}) is much lighter than that of any necessarily-heavy meson that could be exchanged between two subsystems within the tetraquark composite.  It is thus natural to suppose that the favoured structural configuration for a fully-heavy tetraquark state is diquark$+$antidiquark.
Fully-heavy tetraquark states have been studied by some theorists since the 80s \cite{Chao:1980dv,Ader:1981db,Heller:1985cb,Badalian:1985es,Zouzou:1986qh}, even though those investigations did not receive much attention due to the lack of experimental data.

We compute the spectra of $cc\bar c  \bar c$,  $c\bar c b\bar b$, $\bar c \bar c b b \sim c c \bar b \bar b$, $b b \bar b \bar b$ tetraquarks from the diquark$+$antidiquark perspective, using a potential model characterised by linear confinement and one-gluon exchange.  
Using the same approach and assuming isospin symmetry, we also calculate the ground-state masses of similarly viewed $b q \bar b \bar q$, $b b \bar q \bar q$ systems ($q=u,s$).

Finally, as a test of the diquark model approach, we compute the masses of fully-heavy baryons in the diquark model. Our results may be compared soon to the forthcoming experimental data for fully-heavy baryons.

\section{Relativized Diquark Model}
\label{TQ-Model}
We assume that the tetraquark states are colour-antitriplet ($\bar 3_c$) diquark + colour-triplet ($3_c$) antidiquark ($D \bar D$) states.  Furthermore, the constituent $D$, $\bar D$ are each treated as being inert against internal spatial excitations \cite{Anselmino:1992vg, Santopinto:2004hw, Ferretti:2011zz, Santopinto:2014opa}.  This should be a fair approximation for fully-heavy systems owing to the suppression of quark exchange between the diquark subclusters in this case  \cite{Yin:2019bxe}.  Consequently, dynamics within the $D \bar D$ system can be described by a single relative coordinate $\bf{r}_{\rm rel}$, with conjugate momentum ${\bf q}_{\rm rel}$.

To describe the internal dynamics of a $D_a\bar D_b$ system, we choose the Hamiltonian constrained elsewhere \cite{Anwar:2017toa, Anwar:2018sol}:
\begin{subequations}
   \label{eqn:Hmodel}	
\begin{align}
   	\mathcal{H}^{\rm REL} & = T + V({\bf r}_{\rm rel})\,,\\
   T & = \sqrt{{\mathbf q}_{\rm rel}^2+m_{D_a}^2} + \sqrt{{\mathbf q}_{\rm rel}^2+m_{\bar D_b}^2},
\end{align}
\end{subequations}
with the interaction being the sum of a linear-confinement term and a one-gluon exchange (OGE) potential: \cite{Celmaster:1977vh, Godfrey:1985xj, Capstick:1986bm, Anwar:2018sol},
\begin{equation}
	\label{eqn:Vr12-new}  
	\begin{array}{rcl}
		V(r_{\rm rel}) & = & - \frac{3}{4} \Big[\beta r_{\rm rel} + G(r_{\rm rel}) +
    \frac{2 {\bf S}_{D_a} \cdot {\bf S}_{\bar D_b}}{3 m_{D_a} m_{{\bar D}_b}}
		\mbox{ } \nabla^2 G(r_{\rm rel}) \\
		& &- \frac{1}{3 m_{D_a} m_{\bar D_b}} \left(3 {\bf S}_{D_a} \cdot \hat r_{\rm rel} \mbox{ } {\bf S}_{\bar D_b} \cdot \hat r_{\rm rel}
		- {\bf S}_{D_a} \cdot {\bf S}_{\bar D_b}\right) \\
		&  & \times \left(\frac{\partial^2}{\partial r_{\rm rel}^2}
		- \frac{1}{r_{\rm rel}} \frac{\partial}{\partial r_{\rm rel}}\right) G(r_{\rm rel}) + \Delta E \Big] 
		{\bf F}_{D_a} \cdot {\bf F}_{\bar D_b} \mbox{ };
	\end{array}
\end{equation}
it is worth to note that this is the same as \cite[Eq. (7a)]{Anwar:2018sol}; here, we have only included the product of color matrices, ${\bf F}_{D_a}$ and ${\bf F}_{\bar D_b}$, explicitly.\footnote{The {\bf F} symbols of Eq. (\ref{eqn:Vr12-new}) are defined as ${\bf F}_{a,b} = \frac{1}{2} \boldsymbol{\lambda}_{a,b}$, where the $\boldsymbol{\lambda}$s are Gell-Mann color matrices; the underscripts $a$ and $b$ denote (quark/antiquark or diquark/antidiquark) constituents. We also have that $\left\langle {\bf F}_a \cdot {\bf F}_b \right\rangle = \left\{ \begin{array}{ll} - \frac{4}{3} & \mbox{if the constituents $a$ and $b$ are a $q \bar q$, $q D$ or $D \bar D$ pair} \\ - \frac{2}{3} & \mbox{for a $q q$ pair} \end{array} \right.$, where the quarks/diquarks we consider here are color triplets/anti-triplets.}
The Coulomb-like piece is \cite{Godfrey:1985xj, Capstick:1986bm}
\begin{equation}
	\label{eqn:G(r)}
	G(r_{\rm rel}) = - \frac{4 \alpha_{\rm s}(r_{\rm rel})}{3 r_{\rm rel}} =
	- \sum_k \frac{4 \alpha_k}{3 r_{\rm rel}} \mbox{ Erf}(\tau_{ D_a \bar D_b \, k} r_{\rm rel})  \mbox{ }.
\end{equation}
Here, Erf is the error function and \cite{Godfrey:1985xj,Capstick:1986bm}:
\begin{subequations}
\label{eqn:sigma-DD}
\begin{align}
\tau_{ D_a \bar D_b\, k} & = \frac{\gamma_k \sigma_{D_a \bar D_b}}{\sqrt{\sigma_{D_a \bar D_b}^2+\gamma_k^2}} \mbox{ },\\
	\sigma_{D_a \bar D_b}^2 & = \frac{1}{2} \sigma_0^2 \left[1 + \left(\frac{4 m_{D_a} m_{\bar D_b}}{(m_{D_a} + m_{\bar D_b})^2}\right)^4\right] \\
	& + s^2 \left(\frac{2 m_{D_a} m_{\bar D_b}}{m_{D_a} + m_{\bar D_b}}\right)^2
	\mbox{ }.
\end{align}
\end{subequations}
\begin{table}
\begin{ruledtabular}
\begin{tabular*}
{\hsize}
{|l@{\extracolsep{0ptplus1fil}}
l@{\extracolsep{0ptplus1fil}}
|l@{\extracolsep{0ptplus1fil}}
l|@{\extracolsep{0ptplus1fil}}}
Parameter  & Value   & Parameter  & Value \\
\hline
$\alpha_1$ & 0.25                   & $\gamma_1$ & 2.53 fm$^{-1}$  \\
$\alpha_2$ & 0.15                   & $\gamma_2$ & 8.01 fm$^{-1}$  \\
$\alpha_3$ & 0.20                   & $\gamma_3$ & 80.1 fm$^{-1}$  \\
$\sigma_0$ & 9.29 fm$^{-1}$  & $s$ & 1.55  \\
$\beta$        & 3.90 fm$^{-2}$  & $\Delta E$    & $-370$ MeV  \\
$M_{n}$ & $\phantom{5}$220 MeV & $M_{s}$ $\ $& $\phantom{5}$419 MeV  \\
$M_{c}$ & 1628 MeV & $M_{b}$ $\ $& 4977 MeV  \\
$M_{nn}^{\rm sc}$ & $\phantom{5}$691 MeV$\ $ & $M_{nn}^{\rm av}$ & $\phantom{5}$840 MeV$\ $ \\
$M_{ss}^{\rm av}$ & 1136 MeV & & \\
$M_{cs}^{\rm sc}$ & 2229 MeV & $M_{cs}^{\rm av}$ & 2264 MeV \\
$M_{bn}^{\rm sc}$ & 5451 MeV & $M_{bn}^{\rm av}$ $\ $& 5465 MeV  \\
$M_{bs}^{\rm sc}$ & 5572 MeV & $M_{bs}^{\rm av}$ $\ $& 5585 MeV  \\
$M_{bc}^{\rm sc}$ & 6599 MeV & $M_{bc}^{\rm av}$ $\ $& 6611 MeV \\
$M_{cc}^{\rm av}$ & 3329 MeV & $M_{bb}^{\rm av}$ $\ $& 9845 MeV \\
\end{tabular*}
\end{ruledtabular}
\caption{Parameters specifying the Hamiltonian in Eqs.\,\eqref{eqn:Hmodel}.  Here: $n = u$ or $d$; and the superscripts ``sc'' and ``av'' indicate scalar and axial-vector diquarks, respectively. $M_{n,s,c,b}$ are the valence quark masses.}
\label{tab:Model-parameters}
\end{table}
The parameters defining our Hamiltonian are listed in Table~\ref{tab:Model-parameters}.
The strength of the linear confining interaction, $\beta$, and the value of the constant, $\Delta E$, in Eq.\,\eqref{eqn:Vr12-new} are taken from \cite[Table I]{Anwar:2018sol};
and in Eqs.\,\eqref{eqn:G(r)}, \eqref{eqn:sigma-DD}, the values of the parameters $\alpha_k$ and $\gamma_k$ ($k = 1,2,3$), $\sigma_0$ and $s$ are drawn from Refs.\,\cite{Godfrey:1985xj, Capstick:1986bm}.
The valence quark masses are extracted from \cite[Table II]{Godfrey:1985xj}.
This leaves the diquark masses; they are determined by ``binding'' a $(q_1 q_2)^{\rm sc, ax}$ system, $\{q_1, q_2 = n, s, c, b\}$, $n=u=d$, by means of a OGE potential; see \cite{Godfrey:1985xj} and \cite[Table 1]{Ferretti:2019zyh}.
Hence, the results we subsequently report are parameter-free predictions.

If we hypothesize the existence of a dynamical baryon-meson supersymmetry in hadrons \cite{Catto:1984wi,Catto:1987ev}, the Hamiltonian of Eqs. (\ref{eqn:Hmodel}) can also be used to calculate the spectrum of baryons in the quark-diquark approximation.
In the diquark model, baryons and tetraquarks are described as quark-diquark and diquark-antidiquark configurations, respectively, whose internal dynamics is described in terms of a single relative coordinate; see e.g. Refs. \cite{Ferretti:2011zz,Santopinto:2014opa,Santopinto:2004hw}. Because of these premises: I) the solution of either the diquark-antidiquark or quark-diquark eigenvalue problems is substantially the same; II) not only the Hamiltonian of Eqs. (\ref{eqn:Hmodel}) is unchanged, but also the values of the diquark model parameters are expected to remain the same both in the baryon and tetraquark sectors, provided that one substitutes the valence antidiquark, $\bar D$, with a quark, q, with $q = n, s, c$ or $b$.

\section{Fully-heavy baryon and tetraquark states} 
\label{SecResults}
\subsection{$bb\bar q \bar q$ and $b q\bar b \bar q$ ground-state masses}
\label{bbqq}
As an exploratory exercise, we first compute the masses of $J=0^{++}$ heavy-light tetraquarks -- $b q \bar b \bar q$, $b b \bar q \bar q$ systems ($q=n,s$). 

Using the Hamiltonian specified by Eqs.\,\eqref{eqn:Hmodel} and the parameters in Table~\ref{tab:Model-parameters}, we obtain the following ground-state masses:
\begin{equation}
	\label{eqn:Mass-bnbn}
	M_{bn\bar b \bar n}^{\rm gs} = \left\{ \begin{array}{rl} 10.29\,(08) \mbox{ GeV } & (\mbox{sc-sc configuration}) \\
	10.12\,(11) \mbox{ GeV } & (\mbox{av-av configuration}) \end{array} \right.
\end{equation}
and
\begin{equation}
	\label{eqn:Mass-bsbs}
	M_{bs\bar b \bar s}^{\rm gs} = \left\{ \begin{array}{rl} 10.52\,(08) \mbox{ GeV } & (\mbox{sc-sc configuration}) \\
	10.35\,(10) \mbox{ GeV } & (\mbox{av-av configuration}) \end{array} \right.  \mbox{ },
\end{equation}
where the energies of the two possible $D \bar D$ configurations are both shown, viz.\ scalar-scalar and axial-vector--axial-vector.  Evidently, when combining $\bar 3_c$ and $3_c$ constituents, the OGE colour-hyperfine interaction favours a lighter av-av combination.  This is because the spin-spin interaction in Eq.\,\eqref{eqn:Vr12-new} is attractive.
The nature of our uncertainty estimate is discussed in Sec. \ref{AppA} (first procedure). 

Turning now to $bb\bar q \bar q$ systems, we obtain
\begin{subequations}
\label{bbarqqbar}
\begin{align}
	M_{bb\bar n \bar n}^{\rm gs} & = 10.31\,(17)\, \mbox{GeV},\\
	M_{bb\bar s \bar s}^{\rm gs} & = 10.53\,(16) \,\mbox{GeV}.
\end{align}
\end{subequations}
(Owing to Pauli statistics, only av-av configurations are allowed in these cases.)  

In Sec. \ref{AppA} (second procedure), we suggest an alternative prescription to estimate the theoretical uncertainties on our results. If we make use of it, we get for the ground-state (av-av configuration): $M_{bn\bar b \bar n}^{\rm gs} = 10.12\,(19)$ GeV, $M_{bs\bar b \bar s}^{\rm gs} = 10.35\,(19)$ GeV, $M_{bb\bar n \bar n}^{\rm gs} = 10.31\,(22)$ GeV, and $M_{bb\bar s \bar s}^{\rm gs} = 10.53\,(21)$ GeV.
These errors are slightly larger than those in Eqs. (\ref{eqn:Mass-bnbn}-\ref{bbarqqbar}), even though the order of magnitude remains the same.

\begin{table}
\begin{tabular*}
{\hsize}
{|l@{\extracolsep{0ptplus1fil}}
l@{\extracolsep{0ptplus1fil}}
l@{\extracolsep{0ptplus1fil}}
l@{\extracolsep{0ptplus1fil}}
l|@{\extracolsep{0ptplus1fil}}}\hline
Source
& $cc\bar c\bar c$ & $bb\bar b\bar b$ & $bc \bar b \bar c$ & $bb \bar c \bar c$ \\\hline
Threshold & 5968 & 18797 & 12383  & 12550  \\
\hline
Herein & 5883 & 18748$^\dagger$ & 12374 (12521$^*$)  & 12445  \\
\hline
\cite{Anwar:2017toa} & $\cdots$ & 18720 & $\cdots$ & $\cdots$  \\
\cite{Berezhnoy:2011xn} & 5966 & 18754 & $\cdots$ & $\cdots$  \\
\cite{Chen:2016jxd} & 6440 & 18450 & $\cdots$ & $\cdots$  \\
\cite{Karliner:2016zzc} & 6192 & 18826 & $\cdots$ & $\cdots$ \\
\cite{Wu:2016vtq} & 6035 & 18834  & $\cdots$ & $\cdots$   \\
\cite{Wang:2017jtz} & 5990 & 18840  & $\cdots$ & $\cdots$   \\
\cite{Liu:2019zuc} & 6487 &19322 &13035 (13050$^*$) & 12953\\
\cite{Heupel:2012ua} & $(5300\pm500)$ & $\cdots$ & $\cdots$ & $\cdots$ \\
\cite{Debastiani:2017msn} & 5969 & $\cdots$ & $\cdots$ & $\cdots$ \\
\cite{Chen:2020lgj} & 5973 & $\cdots$ & $\cdots$ & $\cdots$ \\
\hline
\end{tabular*}
\caption{\label{tab:tetraquark_masses}
Row~1: experimental meson-meson thresholds. They are $2 \eta_c$, $2 \eta_b$, $\eta_b \eta_c$ and $B_c \bar B_c$ in the $cc\bar c\bar c$, $bb\bar b\bar b$, $bc \bar b \bar c$ and $bb \bar c \bar c$ cases, respectively.
Row~2: our computed results for the masses of $J=0^{++}$ ground-state tetraquark systems. 
For comparison, the other rows list values obtained elsewhere. Masses are in MeV.  All entries describe av-av configurations, except those marked by an asterisk, which are sc-sc. The entry highlighted by $^\dagger$ was obtained previously \cite{Anwar:2017toa}.}
\end{table}

\subsection{$cc\bar c \bar c$, $bb\bar b \bar b$, $b c\bar b \bar c$, $b b \bar c \bar c$ ground-states}
In $QQ\bar Q \bar Q$ systems treated as colour triplet-antitriplet pair configurations, fermion statistics also precludes a role for scalar diquarks.  Consequently, the ground-state $cc\bar c \bar c$ is an av-av combination; and using Eqs.\,\eqref{eqn:Hmodel} we find
\begin{equation}
	\label{eqn:cccc}
	M_{cc\bar c \bar c}^{\rm gs} = 5.88\,(17)\,{\rm GeV}.
\end{equation}
Using the same framework, the calculated mass of the analogous $b b\bar b \bar b$ system is
\begin{equation}
	\label{eqn:bbbb-GS}
	M_{bb\bar b \bar b}^{\rm gs} = 18.75(07)\,{\rm GeV}.
\end{equation}
Table~\ref{tab:tetraquark_masses} lists our prediction for the mass of these systems alongside a sample of values obtained elsewhere \cite{Anwar:2017toa, Berezhnoy:2011xn, Chen:2016jxd, Karliner:2016zzc, Wu:2016vtq, Wang:2017jtz, Liu:2019zuc, Heupel:2012ua, Debastiani:2017msn, Chen:2020lgj}. We also need to mention: Ref. \cite{Ader:1981db}, where the authors stated that for any confining potential there is no state below the threshold; Ref. \cite{Lloyd:2003yc}, where the authors studied tetraquark states in a potential model calculation. They used a kind of individual-particle wave function, with a removal of the center-of-mass energy, $H_{\rm cm} - \frac{3}{2} \hbar \Omega$; see \cite[Eq. (1)]{Lloyd:2003yc}. $\Omega$ was determined from $c \bar c$ states and, as the four-body system is more diluted than a quark-antiquark one, there is a risk of over-correction; Ref. \cite{Richard:2017vry}, where the authors investigated 4-quark states in a potential model with a chromo-electric interaction. 

Considering the $b c\bar b \bar c$ case, both sc-sc and av-av configurations can exist; and we find
\begin{equation}
	\label{eqn:bcbc-GS}
	M_{bc\bar b \bar c}^{\rm gs} = \left\{ \begin{array}{rl} 12.52\,(08) \mbox{ GeV } & (\mbox{sc-sc configuration}) \\
	12.37\,(09) \mbox{ GeV } & (\mbox{av-av configuration}) \end{array} \right.
	\mbox{ }.
\end{equation}
One can also imagine $J=0^{++}$ $bb \bar c \bar c$ ($\bar b \bar b c c$) configurations.  In this case, only the av-av $(bb)_{\bar 3_c}(\bar c\bar c)_{3_c}$ configuration is possible and its ground-state mass is
\begin{equation}
	\label{eqn:bbcc-GS}
	M_{bb\bar c \bar c}^{\rm gs} = 12.45\,(11)\,\mbox{ GeV }.
\end{equation}
Analogously to what is done at the end of Sec. \ref{bbqq}, we make use of the alternative prescription from Sec. \ref{AppA} (second procedure) to give a second estimate of the theoretical uncertainties. We get: $M_{cc\bar c \bar c}^{\rm gs} = 5.88\,(17)$ GeV, $M_{bb\bar b \bar b}^{\rm gs} = 18.75 (23)$ GeV, $M_{bc\bar b \bar c}^{\rm gs} = 12.37\,(20)$ GeV, $M_{bb\bar c \bar c}^{\rm gs} = 12.45\,(21)$ GeV.
Again, even though the theoretical uncertainties resulting from the application of this second prescription are larger than those from Eqs. (\ref{eqn:cccc}-\ref{eqn:bbcc-GS}), they are still of the same order of magnitude.

\subsection{Estimate of Model Uncertainty}
\label{AppA}
In the following, we discuss two prescriptions to estimate the theoretical errors on our predictions of Eqs. (\ref{eqn:Mass-bnbn}-\ref{eqn:bbcc-GS}).
\\{\bf First procedure.} In order to provide an estimate of our theoretical uncertainty, we also compute tetraquark masses using the Relativized Quark Model (RQM) Hamiltonian introduced in Ref.\,\cite{Godfrey:1985xj}. 
Only a few obvious changes are necessary because this Hamiltonian was also constructed to bind a colour triplet-antitriplet pair into a colour-singlet system.
The theoretical error on our results is then given by the difference between the predictions for the tetraquark masses obtained by using: I) the tetraquark model Hamiltonian of Eqs.\,\eqref{eqn:Hmodel}, with the values of the model parameters reported in Table \ref{tab:Model-parameters}; II) the RQM Hamiltonian \cite{Godfrey:1985xj}, where the masses of the quarks are substituted with those of the diquarks, extracted from Table \ref{tab:Model-parameters} herein.

When forming a $S$-wave system from two axial-vector constituents, the only contribution from spin-dependent interactions in the RQM is that produced by the contact term, $V_{\rm cont}$:
\begin{equation}
	\begin{array}{l}
	\left\langle S_1' \mbox{ } S_2' \mbox{ } S' \right| V_{\rm cont}({\bf r})
	\left| S_1 \mbox{ } S_2 \mbox{ } S \right\rangle
	= \left\langle 1 \mbox{ } 1 \mbox{ } 0 \right| V_{\rm cont}({\bf r})
	\left| 1 \mbox{ } 1 \mbox{ } 0 \right\rangle \\
	\hspace{0.5 cm} \propto \frac{1}{2} \left({\bf S}^2 - {\bf S}_1^2 - {\bf S}_2^2 \right)
	= -2 \mbox{ }.
	\end{array}
\end{equation}
Contrarily, in the case of tensor, $V_{\rm tens}$, and spin-orbit, $V_{\rm so}$, interactions, one obtains the matrix elements
\begin{equation}
	\begin{array}{l}
	\left\langle S' \mbox{ } L' \mbox{ } J' \right| V_{\rm tens}({\bf r})
	\left| S \mbox{ } L \mbox{ } J \right\rangle
	= \left\langle 0 \mbox{ } 0 \mbox{ } 0 \right| V_{\rm tens}({\bf r})
	\left| 0 \mbox{ } 0 \mbox{ } 0 \right\rangle \\
	\hspace{0.5 cm} \propto \left\langle L' \right| Y^{(2)} \left| L \right\rangle
	\propto \left( \begin{array}{ccc} 0 & 2 & 0 \\ 0 & 0 & 0 \end{array}\right) = 0 \mbox{ },
	\end{array}
\end{equation}
where $Y^{(2)}$ is a $L=2$ spherical harmonic \cite{Brown:2006cc}, and
\begin{equation}
	\begin{array}{l}
	\left\langle S' \mbox{ } L' \mbox{ } J' \right| V_{\rm so}({\bf r})
	\left| S \mbox{ } L \mbox{ } J \right\rangle
	= \left\langle 0 \mbox{ } 0 \mbox{ } 0 \right| V_{\rm so}({\bf r})
	\left| 0 \mbox{ } 0 \mbox{ } 0 \right\rangle \\
	\hspace{0.5 cm} \propto \sqrt{L (L+1) (2L+1)} = 0 \mbox{ }.
	\end{array}
\end{equation}
The smearing function coefficient employed in Ref.\,\cite{Godfrey:1985xj}, $\sigma_{C_1 C_2}$, with $C_{1,2}$ denoting the constituents, is the same as that we use, given by Eq. (\ref{eqn:sigma-DD}).

As an illustrative example, consider the fully-$b$ $J=0$ tetraquark.  Our prediction for the ground-state mass is reported in Eq.\,\eqref{eqn:bbbb-GS}.  Using the RQM Hamiltonian and a computed value of $\sigma_{(bb)(\bar b \bar b)} = 77.9$ fm$^{-1}$, one finds
\begin{equation}
E^{\rm gs, RQM}_{bb\bar b \bar b} = 18822\,{\rm MeV}.
\end{equation}
Our mass prediction cannot be judged more accurate than the difference between this result and that in Eq.\,\eqref{eqn:bbbb-GS}, \emph{viz}.\ 74\,MeV.  We therefore list this value as the uncertainty in Eq.\,\eqref{eqn:bbbb-GS}.
\\{\bf Second procedure.} Below, we discuss an alternative prescription to estimate the errors on our diquark model results. 
In this second scheme, the errors are extracted by first providing estimates of the uncertainties on the model parameters in Table \ref{tab:Model-parameters} and then by propagating those theoretical errors on the tetraquark model results.
We start by evaluating the theoretical errors on the diquark masses. 

The diquark masses are computed by binding a quark-quark pair via the Relativized QM Hamiltonian \cite{Ferretti:2019zyh,Godfrey:1985xj}. 
The errors on both the quarkonium and diquark masses can be estimated by comparing the RQM predictions of Refs. \cite{Godfrey:1985xj,Godfrey:2004ya,Barnes:2005pb,Godfrey:2015dva,Godfrey:2016nwn,Godfrey:2015dia} to the existing experimental data for heavy and light quarkonia \cite{Tanabashi:2018oca} and extracting the average mass deviations between them. In the case of heavy and heavy-light mesons, we estimate the deviation to be of the order of $0.5\%$. We thus get: $M_{cc}^{\rm av} = 3329 \pm 17$ MeV, $M_{bn}^{\rm sc} = 5451 \pm 27$ MeV, $M_{bn}^{\rm av} = 5465 \pm 27$ MeV, $M_{bs}^{\rm sc} = 5572 \pm 28$ MeV, $M_{bs}^{\rm av} = 5585 \pm 28$ MeV, $M_{bc}^{\rm sc} = 6599 \pm 33$ MeV, $M_{bc}^{\rm av} = 6611 \pm 33$ MeV and $M_{bb}^{\rm av} = 9845 \pm 49$ MeV. In the case of light diquarks, $M_{nn}^{\rm av}$ and $M_{ss}^{\rm av}$, the previous procedure may underestimate the theoretical uncertainties. We thus assign the $M_{nn}^{\rm av}$ and $M_{ss}^{\rm av}$ masses an error of 17 MeV, which is the same as that on $M_{cc}^{\rm av}$. We have: $M_{nn}^{\rm av} = 840 \pm 17$ MeV and $M_{ss}^{\rm av} = 1136 \pm 17$ MeV.

The error on the linear confining potential parameter, $\beta$, is roughly estimated as the difference between the value used herein and in Refs. \cite{Anwar:2017toa,Anwar:2018sol} and that from Ref. \cite{Godfrey:1985xj}. We get: $\beta = 3.90 \pm 0.72$ fm$^{-2}$. We do the same for the constant $\Delta E$: $\Delta E = - 370 \pm 117$ MeV.
The other model parameters, $\alpha_i$ and $\gamma_i$ (with $i = 1, 2, 3$), $\sigma_0$ and $s$, assume the same values both in the RQM \cite{Godfrey:1985xj} and the diquark model of Refs. \cite{Anwar:2017toa,Anwar:2018sol}. Therefore, in first approximation, we assume that their values have no theoretical error.  
By doing this, we somehow underestimate the theoretical error in this second procedure.

\begin{table*}[t]
\begin{ruledtabular}
\begin{center}
\begin{small}
\begin{tabular}{|ccc||ccc||ccc|}
 & ${\bf b} {\bf b} \bar {\bf c} \bar {\bf c}$ {\bf spectrum} & &  & ${\bf c} {\bf c} \bar {\bf c} \bar {\bf c}$ {\bf spectrum} &  &    & ${\bf b} {\bf b} \bar {\bf b} \bar {\bf b}$ {\bf spectrum} &  \\
\hline
$J^{PC}$   & $N[(S_D,S_{\bar D})S,L]J$ & $E^{\rm th}$ [MeV] & $J^{PC}$   & $N[(S_D,S_{\bar D})S,L]J$ & $E^{\rm th}$ [MeV] & $J^{PC}$   & $N[(S_D,S_{\bar D})S,L]J$ & $E^{\rm th}$ [MeV] \\
\hline
$0^{++}$  &  $1[(1,1)0,0]0$ & 12445 & $0^{++}$    &  $1[(1,1)0,0]0$  & 5883 & $0^{++}$    &  $1[(1,1)0,0]0$  & 18748 \\
$0^{++}$  &  $1[(1,1)2,2]0$ & 13208 & $0^{++}$    &  $2[(1,1)0,0]0$  & 6573 & $0^{++}$    &  $2[(1,1)0,0]0$  & 19335 \\
$0^{++}$    &  $2[(1,1)0,0]0$                & 13017
& $0^{++}$    &  $1[(1,1)2,2]0$  & 6827 & $0^{++}$    &  $1[(1,1)2,2]0$  & 19513 \\
$0^{++}$    &  $2[(1,1)2,2]0$                & 13482
& $0^{++}$    &  $3[(1,1)0,0]0$  & 6948 & $0^{++}$    &  $3[(1,1)0,0]0$  & 19644 \\
$0^{++}$    &  $3[(1,1)0,0]0$                & 13349
& $0^{++}$    &  $2[(1,1)2,2]0$  & 7125 & $0^{++}$    &  $2[(1,1)2,2]0$  & 19760 \\
$0^{++}$   &  $3[(1,1)2,2]0$                & 13712
& $0^{++}$    &  $4[(1,1)0,0]0$  & 7237 & $0^{++}$    &  $4[(1,1)0,0]0$  & 19877 \\
&                   &                              & $0^{++}$    &  $3[(1,1)2,2]0$  & 7380 & $0^{++}$    &  $3[(1,1)2,2]0$  & 19964 \\\hline
$1^{+-}$     &  $1[(1,1)1,0]1$  & 12536 & $1^{+-}$     &  $1[(1,1)1,0]1$  & 6120 & $1^{+-}$     &  $1[(1,1)1,0]1$  & 18828 \\
$1^{+-}$     &  $2[(1,1)1,0]1$  & 13060 & $1^{+-}$     &   $2[(1,1)1,0]1$ & 6669 & $1^{+-}$     &   $2[(1,1)1,0]1$ & 19366 \\
$1^{+-}$     &  $3[(1,1)1,0]1$  & 13381 & $1^{+-}$     &  $1[(1,1)1,2]1$  & 6829 & $1^{+-}$     &  $1[(1,1)1,2]1$  & 19511 \\
$1^{+-}$     &  $1[(1,1)1,2]1$  & 13205 & $1^{+-}$     &   $3[(1,1)1,0]1$ & 7016 & $1^{+-}$     &   $3[(1,1)1,0]1$ & 19665 \\
$1^{+-}$     &  $2[(1,1)1,2]1$  & 13479 & $1^{+-}$     &  $2[(1,1)1,2]1$ & 7128 & $1^{+-}$     &  $2[(1,1)1,2]1$ & 19758 \\
$1^{+-}$     &  $3[(1,1)1,2]1$  & 13709 & $1^{+-}$     &   $4[(1,1)1,0]1$ & 7293 & $1^{+-}$     &   $4[(1,1)1,0]1$ & 19893 \\
      &                        &                           & $1^{+-}$     &  $3[(1,1)1,2]1$ & 7382 & $1^{+-}$     &  $3[(1,1)1,2]1$ & 19962 \\\hline
$1^{--}$      &  $1[(1,1)0,1]1$ & 12967 & $1^{--}$      &   $1[(1,1)0,1]1$ & 6580 & $1^{--}$      &   $1[(1,1)0,1]1$ & 19281 \\
$1^{--}$      &  $2[(1,1)0,1]1$ & 13304 & $1^{--}$      &   $1[(1,1)2,1]1$ & 6584 & $1^{--}$      &   $1[(1,1)2,1]1$ & 19288 \\
$1^{--}$      &  $3[(1,1)0,1]1$ & 13565 & $1^{--}$      &   $2[(1,1)0,1]1$ & 6940 & $1^{--}$      &   $2[(1,1)0,1]1$ & 19597 \\
$1^{--}$      &  $1[(1,1)2,1]1$ & 12977 & $1^{--}$      &   $2[(1,1)2,1]1$ & 6943 & $1^{--}$      &   $2[(1,1)2,1]1$ & 19602 \\
$1^{--}$      &  $2[(1,1)2,1]1$ & 13311 & $1^{--}$      &   $3[(1,1)0,1]1$ & 7226 & $1^{--}$      &   $3[(1,1)0,1]1$ & 19833 \\
$1^{--}$      &  $3[(1,1)2,1]1$ & 13572 & $1^{--}$      &   $3[(1,1)2,1]1$ & 7229 & $1^{--}$      &   $3[(1,1)2,1]1$ & 19837 \\\hline
$0^{-+}$     &  $1[(1,1)1,1]0$ & 12976 & $0^{-+}$    &  $1[(1,1)1,1]0$ & 6596 & $0^{-+}$    &  $1[(1,1)1,1]0$ & 19288 \\
$0^{-+}$     &  $2[(1,1)1,1]0$ & 13311 & $0^{-+}$    &  $2[(1,1)1,1]0$ & 6953 & $0^{-+}$    &  $2[(1,1)1,1]0$ & 19602 \\
$0^{-+}$     &  $3[(1,1)1,1]0$ & 13571 & $0^{-+}$    &  $3[(1,1)1,1]0$ & 7236 & $0^{-+}$    &  $3[(1,1)1,1]0$ & 19837 \\\hline
$1^{++}$    & $1[(1,1)2,2]1$ & 13206 & $1^{++}$    &  $1[(1,1)2,2]1$ & 6827 & $1^{++}$    &  $1[(1,1)2,2]1$ & 19512 \\
$1^{++}$    & $2[(1,1)2,2]1$ & 13480 & $1^{++}$    &  $2[(1,1)2,2]1$ & 7125 & $1^{++}$    &  $2[(1,1)2,2]1$ & 19759 \\
$1^{++}$    & $3[(1,1)2,2]1$ & 13710 & $1^{++}$    &  $3[(1,1)2,2]1$ & 7380 & $1^{++}$    &  $3[(1,1)2,2]1$ & 19963 \\\hline
$2^{++}$    &  $1[(1,1)2,0]2$  & 12614  & $2^{++}$    & $1[(1,1)2,0]2$  & 6246 & $2^{++}$    &  $1[(1,1)2,0]2$  & 18900 \\
$2^{++}$    &  $1[(1,1)0,2]2$  & 13204  & $2^{++}$    & $1[(1,1)2,2]2$  & 6827 & $2^{++}$    &  $1[(1,1)2,2]2$  & 19510 \\
$2^{++}$    &  $1[(1,1)2,2]2$  & 13204  & $2^{++}$    & $1[(1,1)0,2]2$  & 6827 & $2^{++}$    &  $1[(1,1)0,2]2$  & 19510 \\
$2^{++}$    &  $2[(1,1)2,0]2$  & 13101  & $2^{++}$    & $2[(1,1)2,0]2$  & 6739 & $2^{++}$    &  $2[(1,1)2,0]2$  & 19398 \\
$2^{++}$    &  $2[(1,1)0,2]2$  & 13478  & $2^{++}$    & $2[(1,1)2,2]2$  & 7125 & $2^{++}$    &  $2[(1,1)2,2]2$  & 19757 \\
$2^{++}$    &  $2[(1,1)2,2]2$  & 13478  & $2^{++}$    & $2[(1,1)0,2]2$  & 7126 & $2^{++}$    &  $2[(1,1)0,2]2$  & 19758 \\
$2^{++}$    &  $3[(1,1)2,0]2$  & 13412  & $2^{++}$    & $3[(1,1)2,0]2$  & 7071 & $2^{++}$    &  $3[(1,1)2,0]2$  & 19688 \\
$2^{++}$    &  $3[(1,1)0,2]2$  & 13708  & $2^{++}$    & $3[(1,1)2,2]2$  & 7380 & $2^{++}$    &  $3[(1,1)2,2]2$  & 19961 \\
$2^{++}$    &  $3[(1,1)2,2]2$  & 13708  & $2^{++}$    & $3[(1,1)0,2]2$  & 7380 & $2^{++}$    &  $3[(1,1)0,2]2$  & 19962 \\
\end{tabular}
\end{small}
\end{center}
\end{ruledtabular}
\caption{\label{tab:bbbb-cccc-spectrum}
Spectra obtained by solving the eigenvalue problem defined by Eqs.\,(\ref{eqn:Hmodel}).  Following Sec. \ref{AppA} (first procedure), the model uncertainty in each result is $\lesssim 2\,$\%.  The states are labelled thus: $N$ is the radial quantum number ($N=1$ is the ground state); $S_{\rm D}$, $S_{\bar {\rm D}}$ are the spin of the diquark and antidiquark, respectively, coupled to the total spin of the meson, $S$; the latter is coupled to the orbital angular momentum, $L$, to get the total angular momentum of the tetraquark, $J$.  Degenerate states are orthogonal combinations of diquark+antidiquark spin vectors \cite{Anwar:2018sol}.}
\end{table*}

Finally, we can use the model parameters of Table \ref{tab:Model-parameters} with their theoretical errors, see above, to calculate the tetraquark masses with their theoretical uncertainties. For example, in the fully-$b$ tetraquark case of Eq. (\ref{eqn:bbbb-GS}) we have: $M_{bb\bar b \bar b}^{\rm gs} = 18.75\,(23)$ GeV.

\begin{table*}[t]
\begin{ruledtabular}
\begin{center}
\begin{small}
\begin{tabular}{|ccc||ccc||ccc|}
\multicolumn{9}{|c|}{$\bf b c \bar b \bar c$ {\bf spectrum}}\\
\hline
$J^{PC}$   & $N[(S_D,S_{\bar D})S,L]J$ & $E^{\rm th}$ [MeV] & $J^{PC}$   & $N[(S_D,S_{\bar D})S,L]J$ & $E^{\rm th}$ [MeV] & $J^{PC}$   & $N[(S_D,S_{\bar D})S,L]J$ & $E^{\rm th}$ [MeV] \\
$0^{++}$ &  $1[(1,1)0,0]0$ & 12374 & $1^{++}$ & $1[(1,0)1,0]1$ & 12533 & $0^{-+}$ &  $1[(1,0)1,1]0$ & 12922 \\
$0^{++}$ &  $1[(0,0)0,0]0$ & 12521 & $1^{++}$ & $1[(1,0)1,2]1$ & 13154 & $0^{-+}$ &  $1[(1,1)1,1]0$ & 12943 \\
$0^{++}$ &  $1[(1,1)2,2]0$ & 13170 & $1^{++}$ & $1[(1,1)2,2]1$ & 13168 & $0^{-+}$ &  $2[(1,0)1,1]0$ & 13250 \\
$0^{++}$ &  $2[(1,1)0,0]0$ & 12975 & $1^{++}$ & $2[(1,0)1,0]1$ & 13036 & $0^{-+}$ &  $2[(1,1)1,1]0$ & 13269 \\
$0^{++}$ &  $2[(0,0)0,0]0$ & 13024 & $1^{++}$ & $2[(1,0)1,2]1$ & 13418 & $0^{-+}$ &  $3[(1,0)1,1]0$ & 13501 \\
$0^{++}$ &  $2[(1,1)2,2]0$ & 13433 & $1^{++}$ & $2[(1,1)2,2]1$ & 13432 & $0^{-+}$ &  $3[(1,1)1,1]0$ & 13519 \\
$0^{++}$ &  $3[(1,1)0,0]0$ & 13301 & $1^{++}$ & $3[(1,0)1,0]1$ & 13342 & & & \\
$0^{++}$ &  $3[(0,0)0,0]0$ & 13330 & $1^{++}$ & $3[(1,0)1,2]1$ & 13638 & & & \\
$0^{++}$ &  $3[(1,1)2,2]0$ & 13653 & $1^{++}$ & $3[(1,1)2,2]1$ & 13652 & & & \\
\hline
$1^{--}$  & $1[(0,0)0,1]1$  & 12910 & $2^{++}$ & $1[(1,1)2,0]2$ & 12576 & $1^{+-}$ & $1[(1,1)1,0]1$ & 12491 \\
$1^{--}$  & $1[(1,0)1,1]1$  & 12922 & $2^{++}$ & $1[(0,0)0,2]2$ & 13143 & $1^{+-}$ & $1[(1,0)1,0]1$ & 12533 \\
$1^{--}$  & $1[(1,1)0,1]1$  & 12934 & $2^{++}$ & $1[(1,1)2,2]2$ & 13166 & $1^{+-}$ & $1[(1,0)1,2]1$ & 13154 \\
$1^{--}$  & $1[(1,1)2,1]1$  & 12944 & $2^{++}$ & $1[(1,1)0,2]2$ & 13166 & $1^{+-}$ & $1[(1,1)1,2]1$ & 13167 \\
$1^{--}$  & $2[(0,0)0,1]1$  & 13238 & $2^{++}$ & $2[(1,1)2,0]2$ & 13063 & $1^{+-}$ & $2[(1,1)1,0]1$ & 13022 \\
$1^{--}$  & $2[(1,0)1,1]1$  & 13250 & $2^{++}$ & $2[(0,0)0,2]2$ & 13406 & $1^{+-}$ & $2[(1,0)1,0]1$ & 13036 \\
$1^{--}$  & $2[(1,1)0,1]1$  & 13262 & $2^{++}$ & $2[(1,1)2,2]2$ & 13429 & $1^{+-}$ & $2[(1,0)1,2]1$ & 13418 \\
$1^{--}$  & $2[(1,1)2,1]1$  & 13269 & $2^{++}$ & $2[(1,1)0,2]2$ & 13430 & $1^{+-}$ & $2[(1,1)1,2]1$ & 13431 \\
$1^{--}$  & $3[(0,0)0,1]1$  & 13490 & $2^{++}$ & $3[(1,1)2,0]2$ & 13365 & $1^{+-}$ & $3[(1,1)1,0]1$ & 13335 \\
$1^{--}$  & $3[(1,0)1,1]1$  & 13501 & $2^{++}$ & $3[(0,0)0,2]2$ & 13627 & $1^{+-}$ & $3[(1,0)1,0]1$ & 13342 \\
$1^{--}$  & $3[(1,1)0,1]1$  & 13513 & $2^{++}$ & $3[(1,1)2,2]2$ & 13650 & $1^{+-}$ & $3[(1,0)1,2]1$ & 13638 \\
$1^{--}$  & $3[(1,1)2,1]1$  & 13519 & $2^{++}$ & $3[(1,1)0,2]2$ & 13650 & $1^{+-}$ & $3[(1,1)1,2]1$ & 13651 \\
\hline
$0^{--}$ &  $1[(1,0)1,1]0$ & 12922 & & & & & & \\
$0^{--}$ &  $2[(1,0)1,1]0$ & 13250 & & & & & & \\
$0^{--}$ &  $3[(1,0)1,1]0$ & 13501 & & & & & & \\
\end{tabular}
\end{small}
\end{center}
\end{ruledtabular}
\caption{As Table \ref{tab:bbbb-cccc-spectrum}, but for $b c \bar b \bar c$ states.}
\label{tab:bbcc-spectrum}
\end{table*}

\subsection{Complete Tetraquark Spectra}
\label{D-aD-spectrum}
The Hamiltonian in Eqs.\,\eqref{eqn:Hmodel} predicts a rich spectrum; and in Tables~\ref{tab:bbbb-cccc-spectrum}, \ref{tab:bbcc-spectrum} we report the lightest states in the spectra of $cc\bar c \bar c$, $b b \bar b \bar b$, $b c\bar b \bar c$ and $b b \bar c \bar c$ systems.  
These results should serve as useful benchmarks for other analyses, which are necessary in order to identify model-dependent artefacts and develop a perspective on those predictions which might only be weakly sensitive to model details.
Moreover, given that the decay modes of fully-$b$ $J=0^{++}$ tetraquarks may be difficult to access experimentally \cite{Aaij:2018zrb,Esposito:2018cwh,Becchi:2020mjz}, our predictions for orbitally-excited and $J\neq 0$ tetraquarks may serve useful in guiding new experimental searches for fully-heavy four-quark states.
Our predictions for the masses of the fully-charm states can be compared to the recent LHCb experimental results \cite{Aaij:2020fnh}. In particular, it can be noticed that we have several candidates in the $6.2-7.4$ GeV energy range with $0^{++}$, $1^{++}$ and $2^{++}$ quantum numbers.

\subsection{$\eta_c$, $\eta_b$ and $B_c$ ground-state masses}
\label{ground-state masses}
The Relativized QM \cite{Godfrey:1985xj} and Relativized Diquark Model \cite{Anwar:2017toa, Anwar:2018sol} Hamiltonians can also be used to calculate the masses of fully-heavy mesons, including the $\eta_c$, the $\eta_b$ and the $B_c$ states. These estimates can provide further informations regarding the theoretical uncertainties in our approach.

Starting from the $\eta_c$, we have
\begin{equation}
	\label{eqn:Eta-c}
	M_{\eta_c} = 2.91\,(06)\,{\rm GeV} \mbox{ }. 
\end{equation}
The Hamiltonian that we use here is that of Eqs. (\ref{eqn:Hmodel}) where, according to hadron supersymmetry, we can substitute the valence diquark/antidiquarks with antiquarks/quarks; the model parameters, including the valence quark masses, are reported in Table \ref{tab:Model-parameters}. The theoretical error, 0.06 GeV, is extracted according to the procedure discussed in Sec. \ref{AppA} (first procedure).

Analogously, for the $\eta_b$ and $B_c$ we have
\begin{equation}
	\label{eqn:Eta-b}
	M_{\eta_b} = 9.33\,(07)\,{\rm GeV} \mbox{ }
\end{equation}
and
\begin{equation}
	\label{eqn:B-c}
	M_{B_c} = 6.18\,(08)\,{\rm GeV} \mbox{ }, 
\end{equation}
respectively.

It is worth noting that the theoretical errors on the meson ground-state masses of Eqs. (\ref{eqn:Eta-c}-\ref{eqn:B-c}), ${\mathcal O}$(100 MeV), are of the same order of magnitude as those on the tetraquark energies of Eqs. (\ref{eqn:Mass-bnbn}-\ref{eqn:bbcc-GS}). Therefore, we can state that the theoretical error on our results for both tetraquarks (Secs. \ref{bbqq}-\ref{D-aD-spectrum}) and ordinary mesons (Sec. \ref{ground-state masses}) is ${\mathcal O}$(100 MeV).

\begin{table*}
\begin{ruledtabular}
\centering
\begin{tabular}{|cccc||cccc|}
$J^P$ & $E_{ccc}^{\rm RQM}$ [MeV] & $E_{ccc}^{\rm RDM}$ [MeV] & $E_{ccc}^{\rm LAT}$ [MeV] & $J^P$ & $E_{bbb}^{\rm RQM}$ [MeV] & $E_{bbb}^{\rm RDM}$ [MeV] & $E_{bbb}^{\rm LAT}$ [MeV] \\
\hline
$\frac{1}{2}^+$ & 5438 & 5271 & $5395\pm13$ & $\frac{1}{2}^+$ & 14926 & 14766 & $14938\pm15\pm16$  \\
$\frac{1}{2}^+$ & 5784 & 5591 & & $\frac{1}{2}^+$ & 15211 & 15028 & $14953\pm14\pm17$ \\
$\frac{1}{2}^+$ & 6084 & 5865 & & $\frac{1}{2}^+$ & 15451 & 15248 & \\
\hline
$\frac{3}{2}^+$ & 4732 & 4655 & $4759\pm6$ & $\frac{3}{2}^+$ & 14247 & 14158 & $14371\pm4\pm11$ \\
$\frac{3}{2}^+$ & 5289 & 5169 & $5313\pm31$ & $\frac{3}{2}^+$ & 14784 & 14652 & $14840\pm12\pm14$ \\
$\frac{3}{2}^+$ & 5682 & 5524 & $5426\pm13$ & $\frac{3}{2}^+$ & 15118 & 14956 & $14958\pm14\pm16$ \\
\end{tabular}
\caption{$\Omega_{ccc}$ and $\Omega_{bbb}$ baryon spectra (up to the 2nd radial excitation), obtained by solving the eigenvalue problems of the Relativized QM (RQM) \cite[Eqs. (2)]{Godfrey:1985xj} and relativized Diquark Model (RDM) [Eq. (\ref{eqn:Hmodel}) herein] Hamiltonians. The values of the model parameters are extracted from \cite[Table II]{Godfrey:1985xj} and Table \ref{tab:Model-parameters}, respectively. 
In the diquark model, the quark-diquark states are labelled as $N(S_q,S_D)S, L; J^P$.
Here, $N$ is the radial quantum number; $S_q$, $S_D$ are the spin of the quark and diquark, respectively, coupled to the total spin of the baryon, $S$; the latter is coupled to the orbital angular momentum, $L$, to get the total angular momentum of the baryon, $J^P$, with parity $P$. 
For $J^P = \frac{1}{2}^+$ states, we have $N(\frac{1}{2},1)\frac{3}{2}, 2; \frac{1}{2}^+$; for $J^P = \frac{3}{2}^+$ states, we have $N(\frac{1}{2},1)\frac{3}{2}, 0; \frac{3}{2}^+$, where $N = 1, 2, 3$.
Our results are compared with those of lattice QCD calculations, Ref. \cite{Padmanath:2013zfa} for the $\Omega_{ccc}$ sector and Ref. \cite{Meinel:2012qz} for the $\Omega_{bbb}$ sector.}
\label{tab:Omega-ccc-bbb-spectrum}
\end{ruledtabular}
\end{table*}

\subsection{Fully-heavy baryon masses}
\label{Fully-heavy baryon ground-state masses}
Here, we extend the diquark model results of the previous sections and Refs. \cite{Anwar:2017toa, Anwar:2018sol} to the study of fully-heavy baryons.

We calculate the ground-state masses of the $\Omega_{ccb}$, $\Omega_{bbc}$, $\Omega_{ccc}$ and $\Omega_{bbb}$ fully-heavy baryons and also those of their radial excitations in the quark-diquark picture. 
We do that by making use of both the Hamiltonian of Eqs. (\ref{eqn:Hmodel}), with the model parameters reported in Table \ref{tab:Model-parameters}, and that of the relativized QM of Ref. \cite{Godfrey:1985xj} (where the diquark mass values are extracted from Table \ref{tab:Model-parameters} herein); analogously to the $Q \bar Q$ meson case of Sec. \ref{ground-state masses}, we also need to substitute a valence antidiquark with a quark.
$\Omega_{QQQ}$ baryons are made up of a heavy quark, $Q = c$ or $b$, and an axial-vector diquark, $\{Q,Q\}$. Scalar-diquark configurations, $[Q,Q]$, are only permitted in the $\Omega_{ccb}$ and $\Omega_{bbc}$ cases.

Starting from the $\Omega_{ccc}$ and $\Omega_{bbb}$ ground-states, characterized by $J^P = \frac{3}{2}^+$ quantum numbers, we have
\begin{equation}
	\label{eqn:Omega-ccc}
	M_{ccc}^{\rm gs} = 4.66\,(08)\,{\rm GeV}
\end{equation}
and
\begin{equation}
	\label{eqn:Omega-bbb}
	M_{bbb}^{\rm gs} = 14.16\,(09)\,{\rm GeV};
\end{equation}
these states are described as $S$-wave quark-axial-vector diquark configurations.
The errors on these results are estimated by using the procedure outlined in Sec. \ref{AppA} (first procedure).
As an example, we discuss the calculation of the $\Omega_{ccc}$ mass with its theoretical uncertainty.
If we make use of the Relativized Diquark Model (RDM) \cite{Anwar:2017toa, Anwar:2018sol}, we get $M_{ccc}^{\rm gs} = 4655$ MeV; on the contrary, if we calculate the mass of the $\Omega_{ccc}$ ground-state in the Relativized QM (RQM) \cite{Godfrey:1985xj}, we obtain $M_{ccc}^{\rm gs} = 4732$ MeV. Our result of Eq. (\ref{eqn:Omega-ccc}) is thus given by the RDM prediction, 4.66 GeV, with the theoretical error being the difference between the previous RDM and the RQM results.
Our predictions for the $\frac{1}{2}^+$ and $\frac{3}{2}^+$ $\Omega_{ccc}$ and $\Omega_{bbb}$ and their radial excitations are reported in Table \ref{tab:Omega-ccc-bbb-spectrum}.
It is worth to note that our RQM and RDM results differ by $\mathcal O$(100 MeV); however, the splittings among radial excitations in the two approaches are very similar. For example, those between the first and second $\frac{3}{2}^+$ radial excitations are 304 MeV (RDM) and 334 MeV (RQM).

In the case of $\Omega_{ccb}$ and $\Omega_{bbc}$ ground-states, we get
\begin{equation}
	\label{eqn:Omega-ccb}
	M_{ccb}^{\rm gs} = 7.72\,(07)\,{\rm GeV}
\end{equation}
and
\begin{equation}
	\label{eqn:Omega-bbc}
	M_{bbc}^{\rm gs} = 11.01\,(12)\,{\rm GeV};
\end{equation}
these states are described as $S$-wave quark-axial-vector diquark configurations, $\{c,c\}b$ and $\{b,b\}c$, with $J^P = \frac{1}{2}^+$ quantum numbers.
For their radial excitations, we get
\begin{equation}
	\label{eqn:Omega-ccb}
	M_{ccb}^{\rm re} = 8.26\,(13)\,{\rm GeV}
\end{equation}
and
\begin{equation}
	\label{eqn:Omega-bbc}
	M_{bbc}^{\rm re} = 11.53\,(16)\,{\rm GeV}.
\end{equation}

Our results can be compared to those of previous studies \cite{Padmanath:2013zfa,Meinel:2012qz,Migura:2006ep,Brown:2014ena,Vijande:2015faa,Serafin:2018aih,Yang:2019lsg,Qin:2019hgk}.
The recent experimental observation of doubly-charmed baryons \cite{Aaij:2017ueg,Aaij:2018wzf} will pave the way, hopefully within a few years, for that of fully-heavy three-quark systems.
Once experimental results for the fully-heavy baryons are released, the quality of the match between our predictions and the data will give indications on the validity of the diquark model approach and the possible existence of a baryon-meson dynamical supersymmetry.

\section{Summary and Perspective}
Adopting a perspective in which tetraquarks are viewed as states made up of elementary colour-antitriplet diquarks and colour-triplet antidiquarks and {baryons as two-body quark-diquark configurations and using a well-constrained model Hamiltonian, built with relativistic kinetic energies, a one-gluon exchange potential and linear confinement [Sec.\,\ref{TQ-Model}], we computed the masses of $b \bar b q \bar q$, $bb \bar q \bar q$, $cc\bar c \bar c$, $b b \bar b \bar b$, $b c\bar b  \bar c$, $b b \bar c \bar c$ states and also fully-heavy $QQQ$ baryons ($Q = c$ or $b$) [Sec.\,\ref{SecResults}].
The eigenvalue problems were solved using a numerical variational procedure in concert with harmonic-oscillator trial wave functions.
In each channel, we compared our prediction for the mass of the $J^{PC}=0^{++}$ tetraquark ground-state (including its theoretical error) with those of previous studies. Our results for the masses of $1^{++}$, $2^{++}$, $0^{++}$ four-quark excited states, and so on, may be of help to the experimentalists in their search for a tetraquark signal in those channels which do not involve the decay of the $0^{++}$ ground-state tetraquark.
Our predictions for $QQQ$ states will be of use to the experimentalists as the recent experimental observation of doubly-charmed baryons \cite{Aaij:2017ueg,Aaij:2018wzf} will pave the way, hopefully within a few years, for that of fully-heavy three-quark systems.
Once experimental results for the fully-heavy baryons are released, the quality of the match between our predictions and the data will give indications on the validity of the diquark model approach and the possible emergence of a baryon-meson dynamical supersymmetry.

While there is a large number of studies on the spectroscopy of tetraquark states, there are few investigations regarding their production \cite{DelFabbro:2004ta,Brodsky:2015wza,Carvalho:2015nqf,Fontoura:2019opw,Yang:2019bbb,Cho:2019lxb,Eichten:2017ual} and principal decay modes \cite{Li:2019uch,Becchi:2020mjz,Anwar:2017toa,Esposito:2018cwh,Wu:2016vtq,Xiao:2019spy,Maiani:2017kyi,Ali:2018ifm,Li:2018bkh,Hernandez:2019eox}.
Some progresses have been made.
In this regard, one can notice that there is a big difference between resonances that can spontaneously dissociate into two mesons, and bound states that decay weakly (e.g., $bb \bar c \bar c$) \cite{Li:2019uch} or by internal annihilation (e.g., $bb \bar b \bar b$) \cite{Becchi:2020mjz}.
We also observe that while the $\Upsilon$ is so narrow, $\mathcal O$(50 keV), because of the Zweig rule, the fully-$b$ ground-state tetraquark case is different; see Ref. \cite{Becchi:2020mjz}. Indeed, in this second instance one can make use of a Fierz transformation to re-write the ground-state tetraquark wave function as a linear combination of $(b \bar b)_1 (b \bar b)_1$ and $(b \bar b)_8 (b \bar b)_8$ components \cite[Eq. (7)]{Becchi:2020mjz}, where the subscripts 1 and 8 indicate color-singlet and octet components, respectively.
Then, a color-octet quark-antiquark component, $(b \bar b)_8$, can couple to a gluon (or, in other words, annihilate into one), giving rise to the possibility of hadronic decays characterized by a relatively large, $\mathcal O$(10 MeV), decay width \cite{Becchi:2020mjz}.

\smallskip

\begin{acknowledgments}
We are grateful for constructive comments from
L.~Maiani,
P.~Bicudo,
Z.-F.~Cui,
G.~Eichmann,
A.~Lovato
and
J.~Segovia.
Work supported by:
CONACyT, M\'exico;
INFN Sezione di Genova;
Instituto de F\'isica y Matem\'aticas, Universidad Michoacana de San Nicol\'as de Hidalgo,
Morelia, Michoac\'an 58040, M\'exico;
Jiangsu Province \emph{Hundred Talents Plan for Professionals};
the Academy of Finland, Project no. 320062;
and
US Department of Energy, Office of Nuclear Physics, contract no.~DE-FG-02-91ER-40608.
\end{acknowledgments}

\end{document}